\pdfoutput=1
\documentclass[runningheads]{llncs}

\usepackage[utf8x]{inputenc}
\usepackage[T1]{fontenc}

\usepackage{comment}
\usepackage{csquotes}
\usepackage{caption} % subfigures
\usepackage{subcaption}

\usepackage{amsmath}
\usepackage{amsfonts}
\usepackage{amssymb}
\usepackage{stmaryrd}
\SetSymbolFont{stmry}{bold}{U}{stmry}{m}{n}
\newcommand*{\qeda}{\hfill\ensuremath{\blacksquare}}%
\newcommand*{\qedb}{\hfill\ensuremath{\square}}%

\usepackage{latexsym}
\usepackage{cmll}
\usepackage{xspace}

\usepackage{mathdots}
\usepackage{nicematrix}
\usepackage{rotating}
\usepackage{tikz}
\usepackage{xcolor}
\definecolor{keywordcolor}{rgb}{0.7, 0.1, 0.1} % red
\definecolor{commentcolor}{rgb}{0.4, 0.4, 0.4} % grey
\definecolor{symbolcolor}{rgb}{0.0, 0.1, 0.6}  % blue
\definecolor{sortcolor}{rgb}{0.1, 0.5, 0.1}    % green
\definecolor{aawhite}{rgb}{0.97,0.97,0.97}
\definecolor{awhite}{rgb}{0.90,0.90,0.90}
\definecolor{lgreen}{rgb}{0.94,1.0,0.98}
\definecolor{dgreen}{rgb}{0.0,0.3,0.1}
\definecolor{sgreen}{rgb}{0.0,0.7,0.3}
\definecolor{lgreen}{rgb}{0.94,1.0,0.98}
\definecolor{bgreen}{rgb}{0.00,0.50,0.25}
\definecolor{dblue}{rgb}{0.0,0.1,0.6}
\definecolor{lorange}{rgb}{1, .85, .60}
\definecolor{lblue}{rgb}{.80, .90, .95}
\definecolor{mixed}{rgb}{0.0,0.3,0.3}
\definecolor{dred}{rgb}{0.6,0.2,0.0}
\definecolor{sred}{rgb}{0.7,0.2,0.0}
\definecolor{ddred}{rgb}{0.3,0.1,0.0}
\definecolor{turq}{rgb}{0.28,0.82,0.80}
\definecolor{lyellow}{rgb}{1.00,0.97,0.94}
\definecolor{mygreen}{rgb}{0,0.6,0}
\definecolor{mygray}{rgb}{0.5,0.5,0.5}
\definecolor{mymauve}{rgb}{0.58,0,0.82}
\definecolor{codegreen}{rgb}{0,0.6,0}
\definecolor{codegray}{rgb}{0.5,0.5,0.5}
\definecolor{codepurple}{rgb}{0.58,0,0.82}
\definecolor{backcolour}{rgb}{0.96,0.96,0.96}

\usepackage{graphicx}

\usepackage{hyperref}
\hypersetup{
    colorlinks=true,
    linkcolor=sred,
    filecolor=magenta,
    urlcolor=cyan,
    citecolor=sgreen,
    bookmarks=true,
}

\usepackage{listings}

\usepackage{float}
\floatstyle{boxed}
\restylefloat{table}
\restylefloat{figure}

\newcommand{\bloch}[2]{\Block[draw=white,fill=lblue,line-width=.5mm,rounded-corners]{#1}{#2}} % https://en.wikipedia.org/wiki/Ernest_Bloch

\newcommand{\NN}{\mathbb{N}}
\newcommand{\ZZ}{\mathbb{Z}}

\newcommand{\floor}[1]{\left\lfloor #1 \right\rfloor}

\newcommand{\gl}{\text{\guillemotleft}}
\newcommand{\gr}{\text{\guillemotright}}

\newcommand{\rppcp}{\mathsf{cp}}

\newcommand{\rppcu}{\mathsf{cu}}

\newcommand{\rpprewire}[1]{\lfloor #1 \rceil}
\newcommand{\rppmkpair}{\mathsf{mkpair}}

\newcommand{\RPP}{\textsf{RPP}\xspace}

\newcommand{\PRF}{\textsf{PRF}\xspace}

\newcommand{\CPP}{\textsf{C}\xspace}
\newcommand{\MATHLIB}{\textsf{mathlib}\xspace}
\newcommand{\LEAN}{\textsf{Lean}\xspace}
\newcommand{\COQ}{\textsf{Coq}\xspace}
\newcommand{\LEANFour}{\textsf{Lean 4}\xspace}
\newcommand{\PISA}{\textsf{Pendulum ISA}\xspace}
\newcommand{\JMF}{\textsf{RI}\xspace} % Jacopini Mentrasti
\newcommand{\Janus}{\textsf{Janus}\xspace}
\newcommand{\Matita}{\textsf{Matita}\xspace}
\newcommand{\SRL}{\textsf{SRL}\xspace}

\newcommand{\yb}{\textcolor{blue}{y}}
\newcommand{\yr}{\textcolor{red}{y}}

\begin{document}
%\title{A formal certification that Reversible Primitive Permutations are Primitive-recursive complete}
\title{Certifying algorithms and relevant properties of Reversible Primitive Permutations with \LEAN}
\titlerunning{Certifying \RPP properties with \LEAN}

\author{Giacomo Maletto\inst{1} \and
	    Luca Roversi\inst{2}\orcidID{0000-0002-1871-6109}}

\authorrunning{G. Maletto, L. Roversi}

\institute{
    Università degli Studi di Torino, Dipartimento di Matematica  -- Italy\\
	\email{giacomo.maletto@edu.unito.it}\\
	\and
	Università degli Studi di Torino, Dipartimento di Informatica -- Italy\\
	\email{luca.roversi@unito.it}}

\maketitle
\begin{abstract}
Reversible Primitive Permutations (\RPP) are recursively defined functions designed to model Reversible Computation. We illustrate a proof, fully developed with the proof-assistant \LEAN, certifying that: ``\RPP can encode every Primitive Recursive Function''. Our reworking of the original proof of that statement is conceptually simpler, fixes some bugs, suggests a new more primitive reversible iteration scheme for \RPP, and, in order to keep formalization and semi-automatic proofs simple, led us to identify a single pattern that can generate some useful reversible algorithms in \RPP: Cantor Pairing, Quotient/Reminder of integer division, truncated Square Root. Our \LEAN source code is available for experiments on Reversible Computation whose properties can be certified.
\end{abstract}

%=====================
\section{Introduction}
\label{section:Introduction}
Studies focused on questions posed by Maxwell, regarding the solidity of the principles which Thermodynamics is based on, recognized the fundamental role that Reversible Computation can play to that purpose.

Once identified, it has been apparent that Reversible Computation constitutes the context in which to frame relevant aspects in areas of Computer Science; they can span from reversible hardware design which can offer a greener foot-print, as compared to classical hardware, to unconventional computational models --- we think of quantum or bio-inspired ones, for example ---, passing through parallel computation and the synchronization issues that it rises, or debuggers that help tracing back to the origin of a bug, or the consistent transactions roll-back in data-base management systems, just to name some. The book \cite{perumalla2013chc} is a comprehensive introduction to the subject; the book \cite{DBLP:books/daglib/0025734}, focused on the low-level aspects of Reversible Computation, concerning the realization of reversible hardware, and
\cite{DBLP:series/eatcs/Morita17}, focused on how models of Reversible Computation like Reversible Turing Machines (RTM), and Reversible Cellular Automata (RCA) can be considered universal and how to prove that they enjoy such a property, are complementary to, and integrate \cite{perumalla2013chc}.

This work focuses on the \emph{functional model} \RPP \cite{DBLP:journals/tcs/PaoliniPR20} of Reversible Computation.
\RPP stands for (the class of) Reversible Primitive Permutations, which can be seen as a possible reversible counterpart of \PRF, the class of Primitive Recursive functions \cite{rogers1967theory}.
We recall that \RPP, in analogy with \PRF, is defined as the smallest class built on some given basic reversible functions, closed under suitable composition schemes.
The very functional nature of the elements in \RPP is at the base of reasonably accessible proofs of the following properties:
\begin{itemize}
\item \RPP is \PRF-complete \cite{DBLP:journals/tcs/PaoliniPR20}: for every function $ F \in \PRF $ with arity $ n \in \NN $, both $ m \in \NN $ and \lstinline|f| in \RPP exist such that \lstinline|f| encodes $ F $, i.e.
$ \mbox{\lstinline|f(|}z,\overline{x},\overline{y}\mbox{\lstinline|) = (|}z+F(\overline{x}),\overline{x},\overline{y}\mbox{\lstinline|)|}$, for every $ \overline{x}\in \NN^n $, whenever all the $m$ variables in $ \overline{y} $ are set to the value $ 0 $.
Both $ z $ and the tuple $ \overline{y} $ are \emph{ancillae}. They can be thought of as temporary storage for intermediate computations of the encoding.

\item \RPP can be extended to become Turing-complete \cite{Paolini2018NGC} by means of a minimization scheme analogous to the one that extends \PRF to the Turing-complete class of \emph{Partial} Recursive Functions.

\item According to \cite{MatosRC2020}, \RPP and the reversible programming language \SRL \cite{matos03tcs} are equivalent, so the fix-point problem is undecidable for \RPP as well \cite{2318_1734164MatosPaoliniRoversiTCSICTCS18}.
\end{itemize}

This work is further evidence that expressing Reversible Computation by means of recursively defined computational models like \RPP, \emph{naturally} offers the possibility to certify with \emph{reasonable effort} the correctness, or other interesting properties, of algorithms in \RPP, by means of some proof-assistant, also discovering new algorithms.
We recall that a proof-assistant is an integrated environment to formalize data-types, to implement algorithms on them, to formalize specifications and prove that they hold, increasing algorithms dependability.

%%%%%%%%%%%%%%%%%%%%%%%%%%%%
\paragraph{Contributions.}
We show how to express \RPP and its evaluation mechanism inside the proof-assistant \LEAN \cite{Lean3}. We can certify the correctness of every reversible function of \RPP with respect to a given specification which also means certifying that \RPP is \PRF-complete, the main result in \cite{DBLP:journals/tcs/PaoliniPR20}. In more detail:
\begin{itemize}
    \item we give a strong guarantee that \RPP is \PRF-complete in three macro steps. We exploit that in \LEAN \MATHLIB library, \PRF is proved equivalent to a class of recursive \emph{unary} functions called \lstinline|primrec|. We define a data-type \lstinline|rpp| in \LEAN to represent \RPP. Then, we certify that, for any function \lstinline|f:primrec|, i.e. any unary \lstinline|f| with type \lstinline|primrec| in \LEAN, a function exists with type \lstinline|rpp| that encodes \lstinline|f:primrec|. Apart from fixing some bugs, our proof is fully detailed as compared to \cite{DBLP:journals/tcs/PaoliniPR20}. Moreover it's conceptually and technically simpler;

    \item concerning simplification, it follows from how the elements in \lstinline|primrec| work, and, additionally, it is characterized by the following aspects:
    \begin{itemize}
        \item we define a \emph{new} finite reversible iteration scheme subsuming the reversible iteration schemes in \RPP, and \SRL, but which is more primitive;

        \item we identify an algorithmic pattern which uniquely associates elements of
        $ \mathbb{N}^2$, and $ \mathbb{N}$ by counting steps in specific paths.
        The pattern becomes a reversible element in \lstinline|rpp| once fixed the parameter it depends on. Slightly different parameter instances generate reversible algorithms whose behavior we can certify in \LEAN. They are truncated Square Root, Quotient/Reminder of integer division, and Cantor Pairing \cite{Cantor1878,DBLP:journals/corr/Szudzik17}.
        The original proof in \cite{DBLP:journals/tcs/PaoliniPR20} that \RPP is \PRF-complete relies on Cantor Pairing, used as a stack to keep the representation of a \PRF function as element of \RPP reversible.
        Our proof in \LEAN replaces Cantor Pairing with a reversible representation of functions \lstinline|mkpair|/\lstinline|unpair| that \MATHLIB supplies as isomorphism $ \mathbb{N}\times\mathbb{N} \simeq \mathbb{N} $. The truncated Square Root is the basic ingredient to obtain reversible \lstinline|mkpair|/\lstinline|unpair|.
    \end{itemize}
\end{itemize}

%---------------------
\paragraph{Related work.}
Concerning the formalization in a proof-assistant of the semantics, and its properties, of a formalism for Reversible Computation, we are aware of \cite{paoliniTYPES2015}. By means of the proof-assistant \Matita \cite{Asperti2007}, it certifies that a denotational semantics for the imperative reversible programming language \Janus \cite[Section 8.3.3]{perumalla2013chc} is fully abstract with respect to the operational semantics.

Concerning \emph{functional models} of Reversible Computation, we are aware of \cite{jacopini89tcs} which introduces the class of reversible functions \JMF, which is as expressive as the \emph{Partial} Recursive Functions. So, \JMF is stronger than \RPP, however we see \JMF as less abstract than \RPP for two reasons: (i) the primitive functions of \JMF depend on a given specific binary representation of natural numbers; (ii) unlike \RPP, which we can see as \PRF in a reversible setting, it is not evident to us that \JMF can be considered the natural extension of a total class analogous to \RPP.

%---------------------
\paragraph{Contents.}
This work illustrates the relevant parts of the BSc Thesis \cite{MalettoBSc2021} which comes with \cite{MalettoRPPLEAN2021}, a \LEAN project that certifies properties, and algorithms of \RPP.
Section~\ref{section:Reversible Primitive Permutations} recalls the class \RPP by commenting on the main design aspects that characterize its definition inside \LEAN.
Section~\ref{section:RPP algorithms} defines and proves correct new reversible algorithms central to the proof.
Section~\ref{section:The UPRF-completeness of RPP} recalls the main aspects of \lstinline|primrec|, and illustrates the key steps to port the original \PRF-completeness proof of \RPP to \LEAN.
Section~\ref{section:Conclusion and developments} is about possible developments.

%=====================
\section{Reversible Primitive Permutations (\RPP) }
\label{section:Reversible Primitive Permutations}

\begin{figure}
    \centering
        \begin{lstlisting}[basicstyle=\small]
        inductive rpp : Type
        -- Base functions
        | Id (n : ℕ) : rpp -- Identity
        | Ne : rpp         -- Sign-change
        | Su : rpp         -- Successor
        | Pr : rpp         -- Predecessor
        | Sw : rpp         -- Transposition or Swap
        -- Inductively defined functions
        | Co (f g : rpp) : rpp   -- Series composition
        | Pa (f g : rpp) : rpp   -- Parallel composition
        | It (f : rpp) : rpp     -- Finite iteration
        | If (f g h : rpp) : rpp -- Selection
        infix `‖`  : 55 := Pa -- Notation for the Parallel composition
        infix `;;` : 50 := Co -- Notation for the Series composition
        \end{lstlisting}
    \caption{The class \RPP as a data-type \lstinline|rpp| in \LEAN.}
    \label{fig:RPP-LEAN}
\end{figure}

We use the data-type \lstinline|rpp| in Figure~\ref{fig:RPP-LEAN}, as defined in \LEAN, to recall from \cite{DBLP:journals/tcs/PaoliniPR20} that the class \RPP is the smallest class of functions
that contains five base functions, named as in the definition, and all the functions that we can generate by the composition schemes whose name is next to the corresponding clause in Figure~\ref{fig:RPP-LEAN}. For ease of use and readability the last two lines in Figure~\ref{fig:RPP-LEAN} introduce infix notations for series and parallel compositions.

\begin{example}[A  term of type {\normalfont \lstinline|rpp|}]
\label{example:A first legal term of type RPP}
In \lstinline|rpp| we can write \lstinline|(Id 1‖Sw);;(It Su)‖(Id 1);;(Id 1‖If Su (Id 1) Pr)| which we also represent as a diagram. Its inputs are the names to the left of the blocks. The outputs are to their right:
\scalebox{}{.85}{
\[
\begin{NiceMatrix}
x&\bloch{1-1}{\mbox{\lstinline|Id 1|}}&x&\bloch{2-1}{\mbox{\lstinline|It  Su|}}&x&\bloch{1-1}{\mbox{\lstinline|Id 1|}}&x
\\
y&\bloch{2-1}{\mbox{\lstinline|Sw|}}&z& &z+x&\bloch{2-1}{\mbox{\lstinline|If Su (Id 1) Pr|}}&z+x
\\
z&&y&\bloch{1-1}{\mbox{\lstinline|Id 1|}}&y&& w
\end{NiceMatrix}
\qquad
\begin{aligned}
w & = \begin{cases}
 y+1 & \textrm{if}\ z+x>0 \\
   y & \textrm{if}\ z+x=0 \\
 y-1 & \textrm{if}\ z+x<0
\end{cases}
\enspace .
\end{aligned}
\]}
We have just built a series composition of three parallel compositions.
The first one composes a unary identity \lstinline|Id 1|, which leaves its unique input untouched, and \lstinline|Sw|, which swaps its two arguments. Then, the $ x $-times iteration of the successor \lstinline|Su|, i.e. \lstinline|It Su|, is in parallel with \lstinline|Id 1|: that is why, one of the outputs of \lstinline|It Su| is $ z + x $. Finally, \lstinline|If Su (Id 1) Pr| selects which among \lstinline|Su|, \lstinline|Id 1|, and \lstinline|Pr| to apply to the argument $ y $, depending on the value of $ z+x $; in particular, \lstinline|Pr| is the function that computes the predecessor of the argument. Figure~\ref{fig:RPP-ev} will give the operational semantics which defines \lstinline|rpp| formally as a class of functions on $ \ZZ$, not on $ \NN $.
\qeda
\end{example}

\begin{remark}[``Weak weakening'' of algorithms in {\normalfont \lstinline|rpp|}]
\label{remark:Weakening algorithms of RPP}
%We can certify that \lstinline|(f‖(Id m)) X = f X|, but not
%\lstinline|((Id m)‖f) X = f X|, for any $ \lstinline|X| $.
We typically drop \lstinline|Id m| if it is the last function of a parallel composition. For example, term and diagram in \textit{Example}~\ref{example:A first legal term of type RPP} become \lstinline|(Id 1‖Sw);;(It Su);;(Id 1‖If Su (Id 1) Pr)| and:
\[ \begin{NiceMatrix}
        x&\bloch{1-1}{\mbox{\lstinline|Id 1|}}&x&\bloch{2-1}{\mbox{\lstinline|It  Su|}}&x&\bloch{1-1}{\mbox{\lstinline|Id 1|}}&x
        \\
        y&\bloch{2-1}{\mbox{\lstinline|Sw|}}&z& &z+x&\bloch{2-1}{\mbox{\lstinline|If Su (Id 1) Pr|}}&z+x
        \\
        z&&y&&y&& w
    \end{NiceMatrix}
    \qquad
    \begin{aligned}
        w & = \begin{cases}
            y+1 & \textrm{if}\ z+x>0 \\
            y & \textrm{if}\ z+x=0 \\
            y-1 & \textrm{if}\ z+x<0
        \end{cases}
        \enspace .
    \end{aligned}
\]
\textit{Remark}~\ref{remark:We keep the definition of ev simple} explains why.
\qeda
\end{remark}

\begin{figure}
\centering
\begin{lstlisting}[basicstyle=\small]
      def arity : rpp → ℕ
        | (Id n)     := n
        | Ne         := 1
        | Su         := 1
        | Pr         := 1
        | Sw         := 2
        | (f ‖ g)    := f.arity + g.arity
        | (f ;; g)   := max f.arity g.arity
        | (It f)     := 1 + f.arity -- It f has an extra argument compared to f
        | (If f g h) := 1 + max (max f.arity g.arity) h.arity
\end{lstlisting}
\caption{Arity of every \lstinline|f: rpp|.}
\label{fig:RPP-arity}
\end{figure}
\noindent
The function in Figure~\ref{fig:RPP-arity} computes the arity of any \lstinline|f:rpp| from the structure of \lstinline|f|, once fixed the arities of the base functions; \lstinline|f.arity| is \LEAN dialect for the more typical notation ``\lstinline|arity(f)|''.

\begin{figure}
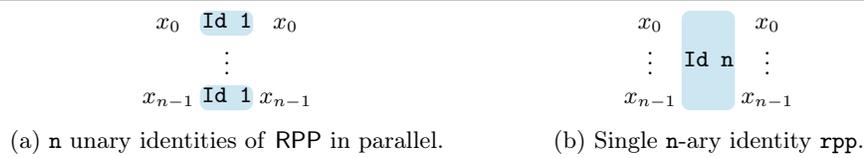

    \begin{subfigure}{.5\textwidth}
        \centering
        $\begin{NiceMatrix}
            x_0&\bloch{1-1}{\mbox{\lstinline|Id 1|}}&x_0
            \\
            &\vdots&
            \\
            x_{n-1}&\bloch{1-1}{\mbox{\lstinline|Id 1|}}&x_{n-1}
        \end{NiceMatrix}$
        \caption{\lstinline|n| unary identities of \RPP in parallel.}
        \label{fig:Id 1 || .. || Id 1}
    \end{subfigure}
    \hfill
    \begin{subfigure}{.45\textwidth}
        \centering
        $ \begin{NiceMatrix}
            x_0&\bloch{3-1}{\mbox{\lstinline|Id n|}}&x_0
            \\
            \vdots&&\vdots
            \\
            x_{n-1}&&x_{n-1}
        \end{NiceMatrix} $
        \caption{Single \lstinline|n|-ary identity \lstinline|rpp|.}
        \label{fig:Id n}
    \end{subfigure}
    \caption{\lstinline|n|-ary identities are base functions of \lstinline|rpp|.}
    \label{fig:multiple-wires}
\end{figure}
\noindent
Figure~\ref{fig:multiple-wires} remarks that \lstinline|rpp| considers \lstinline|n|-ary identities \lstinline|Id n| as primitive; in \RPP the function \lstinline|Id n| is obtained by parallel composition of \lstinline|n| unary identities.

\begin{figure}
\begin{lstlisting}
    def inv : rpp → rpp
      | (Id n)     := Id n -- self-dual
      | Ne         := Ne   -- self-dual
      | Su         := Pr
      | Pr         := Su
      | Sw         := Sw   -- self-dual
      | (f ‖ g)    := inv f ‖ inv g
      | (f ;; g)   := inv g ;; inv f
      | (It f)     := It (inv f)
      | (If f g h) := If (inv f) (inv g) (inv h)
    notation f `⁻¹` := inv f
\end{lstlisting}
\caption{Inverse \lstinline|inv f| of every \lstinline|f:rpp|.}
\label{fig:RPP-inv}
\end{figure}

For any given \lstinline|f:rpp|, the function \lstinline|inv| in Figure~\ref{fig:RPP-inv} builds an element with type \lstinline|rpp|. The definition of \lstinline|inv| lets the successor \lstinline|Su| be inverse of the predecessor \lstinline|Pr| and lets every other base function be self-dual.
Moreover, the function \lstinline|inv| distributes over finite iteration \lstinline|It|, selection \lstinline|If|, and parallel composition \lstinline|‖|, while it requires to exchange the order of the arguments before distributing over the series composition \lstinline|;;|. The last line with \lstinline|notation| suggests that \lstinline|f⁻¹| is the inverse of \lstinline|f|; we shall prove this fact once given the operational semantics of \lstinline|rpp|.

\begin{figure}
\begin{lstlisting}[basicstyle=\small]
   def ev : rpp → list ℤ → list ℤ
   | (Id n)     X                    := X
   | Ne         (x :: X)             := -x :: X
   | Su         (x :: X)             := (x + 1) :: X
   | Pr         (x :: X)             := (x - 1) :: X
   | Sw         (x :: y :: X)        := y :: x :: X
   | (f ;; g)   X                    := ev g (ev f X)
   | (f ‖ g)    X                    := ev f (take f.arity X) ++ ev g (drop f.arity X)
   | (It f)     (x :: X)             := x :: ((ev f)^[↓x] X)
   | (If f g h) (0 :: X)             := 0 :: ev g X
   | (If f g h) (((n : ℕ) + 1) :: X) := (n + 1) :: ev f X
   | (If f g h) (-[1 + n] :: X)      := -[1 + n] :: ev h X
   | _          X                    := X
   notation `‹` f `›` := ev f
\end{lstlisting}
\caption{Operational semantics of elements in \lstinline|rpp|.}
\label{fig:RPP-ev}
\end{figure}

\paragraph{Operational semantics of {\normalfont \lstinline|rpp|}.}
The function \lstinline|ev| in Figure~\ref{fig:RPP-ev} interprets an element of \lstinline|rpp| as a function from a list of integers to a list of integers. Originally, in \cite{DBLP:journals/tcs/PaoliniPR20}, \RPP is a class of functions with type $ \ZZ^n \rightarrow \ZZ^n $. We use \lstinline|list ℤ| in place of tuples of \lstinline|ℤ| to exploit  \LEAN library \MATHLIB and save a large amount of formalization.

Let us give a look at the clauses in Figure~\ref{fig:RPP-ev}.
\lstinline|Id n| leaves the input list \lstinline|X| untouched.
\lstinline|Ne| ``negates'', i.e. takes the opposite sign of, the head of the list, \lstinline|Su| increments, and \lstinline|Pr| decrements it.
\lstinline|Sw| is the transposition, or swap, that exchanges the first two elements of its argument.
The series composition \lstinline|f;;g| first applies \lstinline|f| and then
\lstinline|g|.
The parallel composition \lstinline|f‖g| splits \lstinline|X| into two parts. The ``topmost'' one \lstinline|(take f.arity X)| has as many elements as the arity of \lstinline|f|; the ``lowermost'' one \lstinline|(drop f.arity X)| contains the part of \lstinline|X| that can supply the arguments to \lstinline|g|. Finally, it concatenates the two resulting lists by the append \lstinline|++|. Our \emph{new finite iteration}
\lstinline|It f| iterates \lstinline|f| as many times as the value of the head \lstinline|x| of the argument, if \lstinline|x| contains a non negative value; otherwise it is the identity on the whole \lstinline|x::X|. This behavior is the meaning of \lstinline|(ev f)^[↓x]|.
The selection \lstinline|If f g h| chooses one among \lstinline|f|, \lstinline|g|, and \lstinline|h|, depending on the argument head \lstinline|x|: it is \lstinline|g| with \lstinline|x = 0|, it is \lstinline|f| with \lstinline|x > 0|, and \lstinline|h| with \lstinline|x < 0|.
The last line of Figure~\ref{fig:RPP-ev} sets a handy notation for \lstinline|ev|.

\begin{remark}[We keep the definition of {\normalfont \lstinline|ev|} simple]
\label{remark:We keep the definition of ev simple}
Based on our definition, we can apply any \lstinline|f:rpp| to any \lstinline|X:list ℤ|.
This is based on two observations: first, in \LEAN it holds:
\begin{lstlisting}
    theorem ev_split (f: rpp) (X: list ℤ):
    ‹f› X = (‹f› (take f.arity X)) ++ drop f.arity X
\end{lstlisting}
so that if \lstinline|X.length >= f.arity|, i.e. \lstinline|X| supplies enough arguments, then \lstinline|f| operates on the first elements of \lstinline|X| according to its arity. This justifies \textit{Remark}~\ref{remark:Weakening algorithms of RPP}.
Second, if instead \lstinline|X.length < f.arity| holds, i.e. \lstinline|X| has not enough elements, \lstinline|f X| has an unspecified behavior; this might sound odd, but it simplifies the certified proofs of must-have properties of \lstinline|rpp|.
%This is possible because we can prove:
%\begin{lstlisting}
%    lemma ev_length (f: RPP) (X: list ℤ): (‹f› X).length = X.length
%    theorem ev_split (f: RPP) (X: list ℤ):
%    ‹f› X = ‹f›(take f.arity X) ++ drop f.arity X .
%\end{lstlisting}
%\lstinline|lemma ev_length| says that \lstinline|rpp| functions preserve the argument length.
%One of the most complex property to prove correct, i.e. \lstinline|theorem ev_split|, says that we can apply any \lstinline|‹f›| to \emph{any} \lstinline|X| with at least as many elements as \lstinline|arity f|.
\qeda
\end{remark}

%======================
\subsection{The functions {\normalfont \lstinline|inv h|} and {\normalfont \lstinline|h|} are each other inverse}
Once defined \lstinline|inv| in Figure~\ref{fig:RPP-inv} and \lstinline|ev| in Figure~\ref{fig:RPP-ev} we can prove:
\begin{lstlisting}
    theorem inv_co_l (h : rpp) (X : list ℤ) : ‹h ;; h⁻¹› X = X
    theorem inv_co_r (h : rpp) (X : list ℤ) : ‹h⁻¹ ;; h› X = X
\end{lstlisting}
certifying that \lstinline|h| and \lstinline|h⁻¹| are each other inverse.
We start by focusing on the main details to prove \lstinline|theorem inv_co_l| in \LEAN. The proof proceeds by (structural) induction on \lstinline|h|, which generates 9 cases, one for each clause that defines \lstinline|rpp|. One can go through the majority of them smoothly.
Some comments about two of the more challenging cases follow.

\paragraph{Parallel composition.} Let \lstinline|h| be some parallel composition, whose main constructor is \lstinline|Pa|. The step-wise proof of \lstinline|inv_co_l| is:
\begin{lstlisting}
‹f‖g;;(f‖g)⁻¹› X
    = ‹f‖g;;f⁻¹‖g⁻¹› X       -- by definition
(!) = ‹(f;;f⁻¹)‖(g;;g⁻¹)› X  -- lemma pa_co_pa, arity_inv below
    = ‹f;;f⁻¹›(take f.arity X) ++ ‹g;;g⁻¹›(drop f.arity X)
                             -- by definition
    = take f.arity X ++ drop f.arity X -- by ind. hyp.
    = X                      -- property of ++ (append),
\end{lstlisting}
where the equivalence \lstinline|(!)| holds because we can prove both:
\begin{lstlisting}
lemma pa_co_pa (f f' g g' : rpp) (X : list ℤ) :
  f.arity = f'.arity → ‹f‖g ;; f'‖g'› X = ‹(f;;f') ‖ (g;;g')› X ,
lemma arity_inv (f : rpp) : f⁻¹.arity = f.arity .
\end{lstlisting}
Proving \lstinline|lemma arity_inv|, i.e. that the arity of a function does not change if we invert it, assures that we can prove \lstinline|lemma pa_co_pa|, i.e. that series and parallel compositions smoothly distribute reciprocally.
\qedb

\paragraph{Iteration.} Let \lstinline|h| be a finite iterator whose main constructor is \lstinline|It|.
The goal to prove is \lstinline|‹It f;;It f⁻¹› x::X = x::X| which reduces to \lstinline|‹f⁻¹›^[↓x] (‹f›^[↓x] X') = X'|, where, we recall, the notation \lstinline|‹f›^[↓x]| means ``\lstinline|‹f›| applied \lstinline|x| times, if \lstinline|x| is positive''. Luckily this last statement is both formalized as \lstinline|function.left_inverse g^[n] f^[n]|, available in the library \MATHLIB of \LEAN.
%, where \lstinline|function.left_inverse g^[n] f^[n]| is the proposition \lstinline|∀ (x: α), g(f x) = x : Prop|, with \lstinline|α|, and \lstinline|β| generic types.
\qedb

\vspace{\baselineskip}
To conclude, let us see how the proof of \lstinline|inv_co_r| works. It does not copy-cat the one of \lstinline|inv_co_l|. It relies on proving:
\begin{lstlisting}
   lemma inv_involute (f : rpp) : (f⁻¹)⁻¹ = f ,
\end{lstlisting}
\noindent
which says that applying \lstinline|inv| twice is the identity, and on using \lstinline|inv_co_l|:
\begin{lstlisting}
    ‹f⁻¹ ;; f› X = X -- which, by inv_involute, is equivalent to
    ‹f⁻¹ ;; (f⁻¹)⁻¹› X = X -- which holds because it is an instance of (inv_co_l f⁻¹) .
\end{lstlisting}

\begin{remark}[On our simplifying choices on {\normalfont \lstinline|ev|}]
\label{remark:On our simplifying choices on ev}
A less general, but semantically more appropriate version of  \lstinline|inv_co_l| and \lstinline|inv_co_r| could be:
\begin{lstlisting}
    theorem inv_co_l (f : rpp) (X : list ℤ) :
                f.arity ≤ X.length → ‹f ;; f⁻¹› X = X
    theorem inv_co_r (f : rpp) (X : list ℤ) :
                f.arity ≤ X.length → ‹f⁻¹ ;; f› X = X
\end{lstlisting}
because, recalling \textit{Remark}~\ref{remark:We keep the definition of ev simple}, \lstinline|f X| makes sense when \lstinline|f.arity ≤ X.length|.
Fortunately, the way we defined \lstinline|rpp| allows us to state \lstinline|inv_co_l| or \lstinline|inv_co_r| in full generality
with no reference to \lstinline|f.arity ≤ X.length|.
\qeda
\end{remark}

%======================
\subsection{How {\normalfont \lstinline|rpp|} differs from original {\normalfont \RPP}}
The definition of \lstinline|rpp| in \LEAN is really very close to the original \RPP, but not identical. The goal is to simplify the overall task of formalization and certification. The brief list of changes follows.
\begin{itemize}
    \item As already outlined, \lstinline|It| and \lstinline|If| use the head of the input list to iterate or choose: taking the head of a list with pattern matching is obvious. In \cite{DBLP:journals/tcs/PaoliniPR20}, the last element in the input tuple drives iteration and selection of \RPP.

    \item \lstinline|Id n|, for any \lstinline|n:ℕ|, is primitive in \lstinline|rpp| and derived in \RPP.
%    In some cases it is useful to have \lstinline|Id 0| available.

    \item Using \lstinline|list ℤ → list ℤ| as the domain of the function that interprets any given element \lstinline|f:rpp| avoids  letting the type of \lstinline|f:rpp| depend on the arity of \lstinline|f|. To know the arity of \lstinline|f| it is enough to invoke \lstinline|arity f|. Finally, we observe that getting rid of a dependent type like, say, \lstinline|rpp n|, allows us to escape situations in which we would need to compare equal but not definitionally equal types like \lstinline|rpp (n+1)| and \lstinline|rpp (1+n)|.

    \item The new finite iterator \lstinline|It f (x::t): list ℤ| \emph{subsumes} the finite iterators \lstinline|ItR| in \RPP, and \lstinline|for| in \SRL, i.e. \lstinline|It| is more primitive, equally expressive and
    simpler for \LEAN to prove that its definition is terminating.

    We recall that \lstinline|ItR f (x₀,x₁,...,xₙ₋₂,x)| evaluates to \lstinline|f(f(...f(x₀,x₁,...,xₙ₋₂)...))| with $ \mid\!\!\mbox{\lstinline|x|}\!\!\mid $ occurrences of \lstinline|f|.
    Instead, \lstinline|for(f) x| evaluates to \lstinline|f(f(...f(x₀,x₁,...,xₙ₋₂)...))|, with \lstinline|x| occurrences of \lstinline|f|, if \lstinline|x > 0|; it evaluates to \lstinline|f⁻¹(f⁻¹(...f⁻¹(x₀,x₁,...,xₙ₋₂)...))|, with \lstinline|-x| occurrences of \lstinline|f⁻¹|, if \lstinline|x < 0|; it behaves like the identity if \lstinline|x = 0|.

    We can define both \lstinline|ItR| and \lstinline|for| in terms of \lstinline|It|:
    \begin{align}
    \label{align:It ItR}
        \mbox{\lstinline|ItR f|}
        & =
        \mbox{\lstinline|(It f);;Ne;;(It f);;Ne|} \\
    \label{align:It for}
        \mbox{\lstinline|for(f)|}
        & =
        \mbox{\lstinline|(It f);;Ne;;(It f⁻¹);;Ne|}
        \enspace .
    \end{align}
    \begin{example}[How does \eqref{align:It ItR} work?]
    \label{example:How align:It ItR works}
    Whenever \lstinline|x > 0|, the leftmost \lstinline|It f| in \eqref{align:It ItR} iterates \lstinline|f|, while the rightmost one does nothing because \lstinline|Ne| in the middle negates \lstinline|x|.
    On the contrary, if \lstinline|x < 0|, the leftmost \lstinline|It f| does nothing and the iteration is performed by the rightmost iteration, because \lstinline|Ne| in the middle negates \lstinline|x|. In both cases, the last \lstinline|Ne| restores \lstinline|x| to its initial sign. But this is the behavior of \lstinline|ItR|, as we wanted. \qeda
    \end{example}
\end{itemize}

%=====================
\section{\RPP algorithms central to our proofs}
\label{section:RPP algorithms}

\begin{figure}
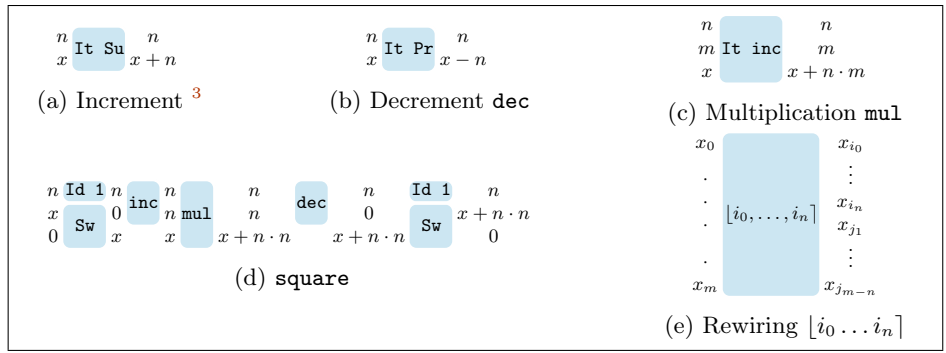

\begin{subfigure}{.225\textwidth}
\centering
\scalebox{.8}
{$\begin{NiceMatrix}
    n & \bloch{2-1}{\mbox{\lstinline|It Su|}} & n     \\
    x &                      & x + n \\
 \end{NiceMatrix}$
}\caption{Increment \footnotemark }
\label{sfig:inc}
\end{subfigure}
\hfill
\begin{subfigure}{.225\textwidth}
\centering
\scalebox{.8}
{$ \begin{NiceMatrix}
    n & \bloch{2-1}{\mbox{\lstinline|It Pr|}} & n     \\
    x &                      & x - n \\
  \end{NiceMatrix} $
}
\caption{Decrement \lstinline|dec|}
\label{sfig:dec}
\end{subfigure}
\hfill
\begin{subfigure}{.325\textwidth}
\centering
\scalebox{.8}
{$ \begin{NiceMatrix}
    n & \bloch{3-1}{\mbox{\lstinline|It inc|}} & n             \\
    m &                     & m             \\
    x &                     & x + n \cdot m \\
  \end{NiceMatrix} $
}\caption{Multiplication \lstinline|mul|}
\label{sfig:mul}
\end{subfigure}
\\
\begin{subfigure}{.6\textwidth}
\centering
\scalebox{.8}
{$\begin{NiceMatrix}
    n &  \bloch{1-1}{\mbox{\lstinline|Id 1|}}  & n & \bloch{2-1}{\mbox{\lstinline|inc|}} & n & \bloch{3-1}{\mbox{\lstinline|mul|}} & n             & \bloch{2-1}{\mbox{\lstinline|dec|}} & n             & \bloch{1-1}{\mbox{\lstinline|Id 1|}}   & n
    \\
    x & \bloch{2-1}{\mbox{\lstinline|Sw|}} & 0 &   & n &                      & n             &                      & 0             & \bloch{2-1}{\mbox{\lstinline|Sw|}} & x + n \cdot n
    \\
    0 &                     & x &   & x &                      & x + n \cdot n &   \      & x + n \cdot n &                     & 0
\end{NiceMatrix}$
}
\caption{\lstinline|square|}
\label{fig:square}
\end{subfigure}
\hfill
\begin{subfigure}{.325\textwidth}
    \centering
\scalebox{.8}
{    $\begin{NiceMatrix}
        x_0 & \bloch{6-1}{\rpprewire{i_0, \ldots, i_n}} & x_{i_0} \\
      \cdot &                                & \vdots  \\
      \cdot &                                & x_{i_n} \\
      \cdot &                                & x_{j_1} \\
      \cdot &                                & \vdots  \\
        x_m &                                & x_{j_{m-n}}
    \end{NiceMatrix}$
}    \caption{Rewiring $ \rpprewire{i_0 \ldots i_n} $}
    \label{fig:rewiring}
\end{subfigure}
\caption{Some useful functions of \lstinline|rpp|}
\label{fig:standard functions}
\end{figure}
\footnotetext{Note that using our definition, the variable \lstinline|n| must be non-negative in order to have the shown behavior,
otherwise the function acts as the identity. This is why it's called {\it increment} and not {\it addition}.}

Figure~\ref{fig:standard functions} recalls definition, and behavior of some \lstinline|rpp| functions in \cite{DBLP:journals/tcs/PaoliniPR20}. It is worth commenting on how rewiring $ \rpprewire{i_0 \ldots i_n} $ works. Let $\{ i_0, \dots, i_n \} \subseteq \{ 0, \dots, m \}$. Let $\{ j_1, \dots, j_{m-n} \}$ be the set of remaining indices $\{ 0, \dots, m \} \setminus \{ i_0, \dots, i_n \}$ ordered such that $j_k < j_{k+1}$.
By definition, $\rpprewire{i_0, \dots, i_n} (x_0, \dots, x_m) = (x_{i_0}, \dots, x_{i_n},$ $x_{j_1}, \dots, x_{j_{m-n}})$, i.e. rewiring brings every input with index in $ \{ i_0, \dots,  i_n \} $ before all the remaining inputs, preserving the order.

\begin{figure}
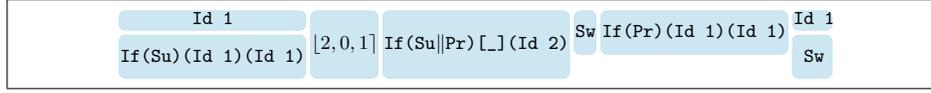

    \centering
    \scalebox{.825}{
        $\begin{NiceMatrix}
            &\bloch{1-1}{\mbox{\lstinline|Id 1|}} & \bloch{3-1}{\rpprewire{2,0,1}} & \bloch{3-1}{\mbox{\lstinline|If(Su‖Pr)[_](Id 2)|}} & \bloch{2-1}{\mbox{\lstinline|Sw|}} & \bloch{2-1}{\mbox{\lstinline|If(Pr)(Id 1)(Id 1)|}} &\bloch{1-1}{\mbox{\lstinline|Id 1|}}\\
            &\bloch{2-1}{\mbox{\lstinline|If(Su)(Id 1)(Id 1)|}}&&     &&    & \bloch{2-1}{\mbox{\lstinline|Sw|}}\\
            &&&&&
        \end{NiceMatrix}$
       }
    \caption{Algorithm scheme \lstinline|step[_]|. The algorithm we can obtain from it depends on how we fill the hole \lstinline|[_]|.}
    \label{fig:function scheme step}
\end{figure}

Figure~\ref{fig:function scheme step} identifies the new algorithm scheme \lstinline|step[_]|. Depending on how we fill the hole \lstinline|[_]|, we get step functions that, once iterated, draw paths in $ \NN^2 $.
%The names $ a_0, \ldots, b_0, \ldots $ are just place holders, currently.

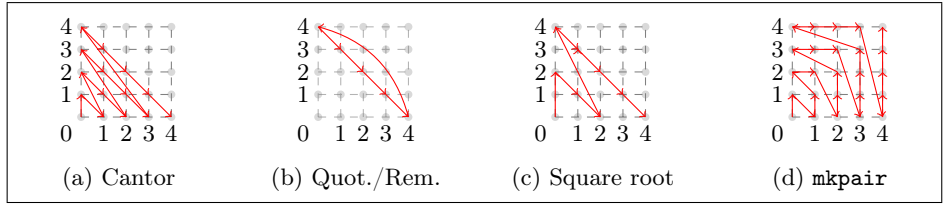
\begin{figure}
\begin{subfigure}{.225\textwidth}
\centering
\begin{tikzpicture}[scale=0.3]
    % grid
    \draw[dashed,fill=gray,opacity=.5]  (0,0) grid (4,4);
    % labels
    \foreach \x in {1,...,4} { \node [anchor=north] at (\x,0) {\x}; }
    \foreach \y in {1,...,4} { \node [anchor=east] at (0,\y) {\y}; }
    \node [anchor=north east] at (0,0) {0};
    % vertices
    \foreach \x in {0,...,4} {
        \foreach \y in {0,...,4} {
            \node at (\x,\y) [circle,inner sep=0pt,minimum size=3pt,fill=gray,opacity=.3] {};
        }
    }
    % path
    \draw [->,red] (0,0) -- (0,1);
    \draw [->,red] (0,1) -- (1,0);
    \draw [->,red] (1,0) -- (0,2);
    \draw [->,red] (0,2) -- (1,1);
    \draw [->,red] (1,1) -- (2,0);
    \draw [->,red] (2,0) -- (0,3);
    \draw [->,red] (0,3) -- (1,2);
    \draw [->,red] (1,2) -- (2,1);
    \draw [->,red] (2,1) -- (3,0);
    \draw [->,red] (3,0) -- (0,4);
    \draw [->,red] (0,4) -- (1,3);
    \draw [->,red] (1,3) -- (2,2);
    \draw [->,red] (2,2) -- (3,1);
    \draw [->,red] (3,1) -- (4,0);
\end{tikzpicture}
\caption{Cantor}
\label{sfig:cantor as a path}
\end{subfigure}
\hfill
\begin{subfigure}{.225\textwidth}
\centering
\begin{tikzpicture}[scale=0.3]
    % grid
    \draw[dashed,fill=gray,opacity=.3]  (0,0) grid (4,4);
    % labels
    \foreach \x in {1,...,4} { \node [anchor=north] at (\x,0) {\x}; }
    \foreach \y in {1,...,4} { \node [anchor=east] at (0,\y) {\y}; }
    \node [anchor=north east] at (0,0) {0};
    % vertices
    \foreach \x in {0,...,4} {
        \foreach \y in {0,...,4} {
            \node at (\x,\y) [circle,inner sep=0pt,minimum size=3pt,fill=gray,opacity=.3] {};
        }
    }
    % path
\draw [->,red] (4,0) to [bend right=30] (0,4);
\draw [->,red] (0,4) -- (1,3);
\draw [->,red] (1,3) -- (2,2);
\draw [->,red] (2,2) -- (3,1);
\draw [->,red] (3,1) -- (4,0);
\end{tikzpicture}
\caption{Quot./Rem.}
\label{sfig:Quot./Rem.}
\end{subfigure}
\hfill
\begin{subfigure}{.225\textwidth}
\centering
\begin{tikzpicture}[scale=0.3]
    % grid
    \draw[dashed,fill=gray,opacity=.5]  (0,0) grid (4,4);
    % labels
    \foreach \x in {1,...,4} { \node [anchor=north] at (\x,0) {\x}; }
    \foreach \y in {1,...,4} { \node [anchor=east] at (0,\y) {\y}; }
    \node [anchor=north east] at (0,0) {0};
    % vertices
    \foreach \x in {0,...,4} {
        \foreach \y in {0,...,4} {
            \node at (\x,\y) [circle,inner sep=0pt,minimum size=3pt,fill=gray,opacity=.3] {};
        }
    }
    % path
\draw [->,red] (0,0) -- (0,2);
\draw [->,red] (0,2) -- (1,1);
\draw [->,red] (1,1) -- (2,0);
\draw [->,red] (2,0) -- (0,4);
\draw [->,red] (0,4) -- (1,3);
\draw [->,red] (1,3) -- (1,3);
\draw [->,red] (1,3) -- (2,2);
\draw [->,red] (2,2) -- (3,1);
\draw [->,red] (3,1) -- (4,0);
\end{tikzpicture}
\caption{Square root}
\label{sfig:Square root}
\end{subfigure}
\hfill
    \begin{subfigure}{.225\textwidth}
    \centering
    \begin{tikzpicture}[scale=0.3]
        % grid
        \draw[dashed,fill=gray,opacity=.5]  (0,0) grid (4,4);
        % labels
        \foreach \x in {1,...,4} { \node [anchor=north] at (\x,0) {\x}; }
        \foreach \y in {1,...,4} { \node [anchor=east] at (0,\y) {\y}; }
        \node [anchor=north east] at (0,0) {0};
        % vertices
        \foreach \x in {0,...,4} {
            \foreach \y in {0,...,4} {
                \node at (\x,\y) [circle,inner sep=0pt,minimum size=3pt,fill=gray,opacity=.3] {};
            }
        }
        % path
        \draw [->,red] (0,0) -- (0,1);
        \draw [->,red] (0,1) -- (1,0);
        \draw [->,red] (1,0) -- (1,1);
        \draw [->,red] (1,1) -- (0,2);
        \draw [->,red] (0,2) -- (1,2);
        \draw [->,red] (1,2) -- (2,0);
        \draw [->,red] (2,0) -- (2,1);
        \draw [->,red] (2,1) -- (2,2);
        \draw [->,red] (2,2) -- (0,3);
        \draw [->,red] (0,3) -- (1,3);
        \draw [->,red] (1,3) -- (2,3);
        \draw [->,red] (2,3) -- (3,0);
        \draw [->,red] (3,0) -- (3,1);
        \draw [->,red] (3,1) -- (3,2);
        \draw [->,red] (3,2) -- (3,3);
        \draw [->,red] (3,3) -- (0,4);
        \draw [->,red] (0,4) -- (1,4);
        \draw [->,red] (1,4) -- (2,4);
        \draw [->,red] (2,4) -- (3,4);
        \draw [->,red] (3,4) -- (4,0);
        \draw [->,red] (4,0) -- (4,1);
        \draw [->,red] (4,1) -- (4,2);
        \draw [->,red] (4,2) -- (4,3);
        \draw [->,red] (4,3) -- (4,4);
    \end{tikzpicture}
    \caption{\lstinline|mkpair|}
    \label{sfig:mkpair}
\end{subfigure}
\caption{Paths in $ \NN^2 $ that generate algorithms in \lstinline|rpp|.}
\label{fig:functions as paths}
\end{figure}

On top of the functions in Figures~\ref{fig:standard functions}, and~\ref{fig:function scheme step} we build Cantor Pairing/Un-pairing, Quotient/Reminder of integer division, and truncated Square Root. It is enough to make the correct instance of \lstinline|step[_]| in order to visit $ \NN^2 $ as in Figures~\ref{sfig:cantor as a path}, \ref{sfig:Quot./Rem.}, and~\ref{sfig:Square root}, respectively. The alternative pairing $\rppmkpair$ has a more complex definition, and is a necessary ingredient for the main proof.

\paragraph{Cantor (Un-)Pairing}. The standard definition of Cantor Pairing $\rppcp : \NN^2 \to \NN$ and Un-pairing $\rppcu : \NN \to \NN^2$, two bijections one inverse of the other, is:
\begin{align}
\label{align: cp analitically}
\rppcp(x,y) & = \sum_{i=1}^{x+y}i + x = \frac{(x + y)(x + y + 1)}{2} + x
\\
\label{align: cu analitically}
\rppcu(n)  & = \left( n - \frac{i(1+i)}{2}, \frac{i(3+i)}{2} - n \right)
\enspace ,
\end{align}
where $i = \left\lfloor \frac{\sqrt{8n + 1} - 1}{2} \right\rfloor $.

\begin{figure}
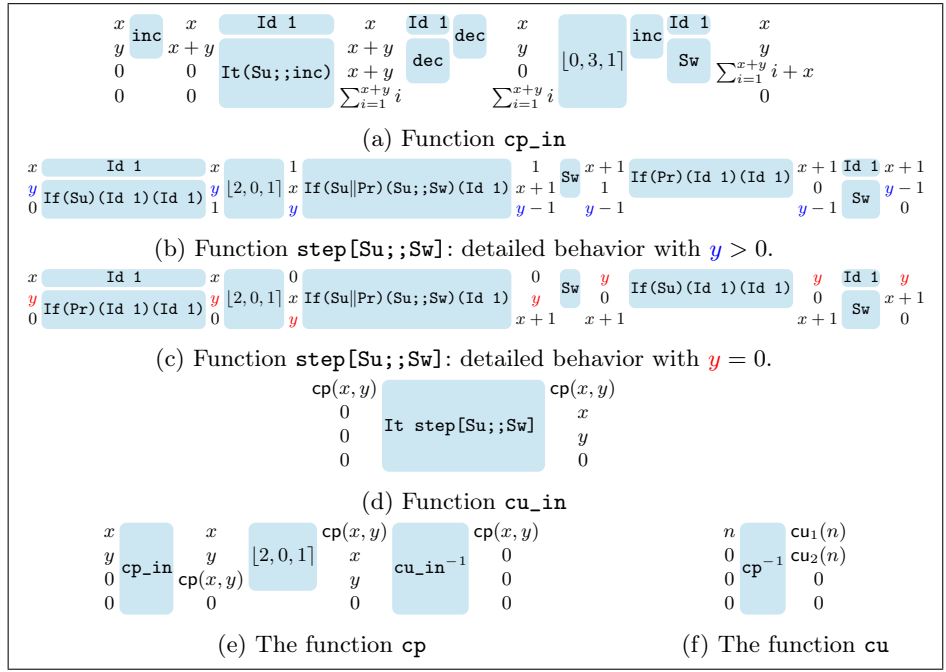

\begin{subfigure}{.975\textwidth}
    \centering
\scalebox{.82}{
    $\begin{NiceMatrix}
        x &\bloch{2-1}{\mbox{\lstinline|inc|}}&x&\bloch{1-1}{\mbox{\lstinline|Id 1|}}&x&\bloch{1-1}{\mbox{\lstinline|Id 1|}}&\bloch{2-1}{\mbox{\lstinline|dec|}}&x& \bloch{4-1}{\rpprewire{0,3,1}} &\bloch{2-1}{\mbox{\lstinline|inc|}} &\bloch{1-1}{\mbox{\lstinline|Id 1|}} & x
        \\
        y && x + y & \bloch{3-1}{
            \mbox{\lstinline|It(Su;;inc)|}
            %            \rppIt[\rppSu \rppCo \rppinc]
        } & x + y& \bloch{2-1}{\mbox{\lstinline|dec|}} && y&&& \bloch{2-1}{\mbox{\lstinline|Sw|}} & y
        \\
        0 &                      & 0     &                     & x + y                &                      &                      & 0                    &                      &                      &                     & \sum_{i=1}^{x + y} i + x
        \\
        0 &                      & 0     &                     & \sum_{i=1}^{x + y} i &                      &                      & \sum_{i=1}^{x + y} i &                      &                      &                     & 0
    \end{NiceMatrix}$
}
    \caption{Function \lstinline|cp_in|}
    \label{sfig:cp input}
\end{subfigure}
\\
\begin{subfigure}{.975\textwidth}
\centering
\scalebox{.72}{
    $\begin{NiceMatrix}
        & x &\bloch{1-1}{\mbox{\lstinline|Id 1|}}& x & \bloch{3-1}{\rpprewire{2,0,1}} & 1 & \bloch{3-1}{
            \mbox{\lstinline|If(Su‖Pr)(Su;;Sw)(Id 1)|}
%            \rppIf[\boldsymbol{\rppSu \rppPa \rppPr}, \mbox{\lstinline|Su;;Sw|}, \ \rppId_1]
        } & 1     & \bloch{2-1}{
        \mbox{\lstinline|Sw|}
%        \rppSw
        } & x + 1 & \bloch{2-1}{
        \mbox{\lstinline|If(Pr)(Id 1)(Id 1)|}
%        \rppIf[\boldsymbol{\rppPr}, \rppId_1, \rppId_1]
        } & x + 1 &\bloch{1-1}{\mbox{\lstinline|Id 1|}} & x + 1 \\
        & \yb & \bloch{2-1}{
        \mbox{\lstinline|If(Su)(Id 1)(Id 1)|}
        %            \rppIf[\boldsymbol{\rppSu}, \rppId_1, \rppId_1]
        } & \yb &&x&& x + 1 &                     & 1     &                                                              & 0     & \bloch{2-1}{
        \mbox{\lstinline|Sw|}
%        \rppSw
        } & \yb - 1 \\
        & 0&&1&&\yb&& \yb - 1 && \yb - 1 && \yb - 1 && 0
    \end{NiceMatrix}$
}
\caption{Function \lstinline|step[Su;;Sw]|: detailed behavior with $ \yb > 0 $. }
\label{sfig:cu step y > 0}
\end{subfigure}
\\
\begin{subfigure}{.975\textwidth}
\centering
\scalebox{.72}{
$\begin{NiceMatrix}
    & x &\bloch{1-1}{\mbox{\lstinline|Id 1|}}& x & \bloch{3-1}{\rpprewire{2,0,1}} & 0 & \bloch{3-1}{
    \mbox{\lstinline|If(Su‖Pr)(Su;;Sw)(Id 1)|}
    %    \rppIf[\rppSu \rppPa \rppPr, \mbox{\lstinline|Su;;Sw|} , \ \rppId_1]
    } & 0     & \bloch{2-1}{
    \mbox{\lstinline|Sw|}
    %    \rppSw
    } & \yr & \bloch{2-1}{
        \mbox{\lstinline|If(Su)(Id 1)(Id 1)|}
%        \rppIf[\rppPr, \boldsymbol{\rppId_1}, \rppId_1]
    } & \yr & \bloch{1-1}{\mbox{\lstinline|Id 1|}} & \yr
    \\
    & \yr & \bloch{2-1}{
        \mbox{\lstinline|If(Pr)(Id 1)(Id 1)|}
%        \rppIf[\rppSu, \boldsymbol{\rppId_1}, \rppId_1]
    } & \yr && x &&\yr&&0&&0&\bloch{2-1}{
        \mbox{\lstinline|Sw|}
%        \rppSw
   } & x + 1
    \\
    & 0 && 0 && \yr && x + 1 && x + 1 && x + 1 && 0
\end{NiceMatrix}$
}
\caption{Function \lstinline|step[Su;;Sw]|: detailed behavior with $ \yr = 0 $. }
\label{sfig:cu step y = 0}
\end{subfigure}
\\
\begin{subfigure}{.975\textwidth}
\centering
\scalebox{.82}{
$\begin{NiceMatrix}
     \rppcp(x,y)
      & \bloch{4-1}{\mbox{\lstinline|It step[Su;;Sw]| }} &
      \rppcp(x,y)\\
    0 &                      & x \\
    0 &                      & y \\
    0 &                      & 0
\end{NiceMatrix}$
}
\caption{Function \lstinline|cu_in|}
\label{sfig:cu input}
\end{subfigure}
\\
\begin{subfigure}{.665\textwidth}
    \centering
\scalebox{.82}{
    $ \begin{NiceMatrix}
        x & \bloch{4-1}{\mbox{\lstinline|cp_in|}} & x           & \bloch{3-1}{\rpprewire{2,0,1}} & \rppcp(x,y) & \bloch{4-1}{\mbox{\lstinline|cu_in⁻¹|}} & \rppcp(x,y) \\
        y &                       & y           &                      & x           &                            & 0           \\
        0 &                       & \rppcp(x,y) &                      & y           &                            & 0           \\
        0 &                       & 0           &                      & 0           &                            & 0           \\
    \end{NiceMatrix}
    $
}
\caption{The function \lstinline|cp|}
\label{sfig:cp}
\end{subfigure}
\hfill
\begin{subfigure}{.325\textwidth}
\centering
\scalebox{.82}{
    $ \begin{NiceMatrix}
        n & \bloch{4-1}{\mbox{\lstinline|cp⁻¹|}} & \rppcu_1(n)    \\
        0 &                      & \rppcu_2(n)         \\
        0 &                      & 0 \\
        0 &                      & 0
    \end{NiceMatrix} $
}
\caption{The function \lstinline|cu|}
\label{sfig:cu}
\end{subfigure}
\caption{Cantor Pairing and Un-pairing.}
\label{fig:Cantor}
\end{figure}

Figure~\ref{fig:Cantor} has all we need to define Cantor Pairing \lstinline|cp:rpp|, and Un-pairing \lstinline|cu:rpp|.
In Figure~\ref{sfig:cp input}, \lstinline|cp_in| is the natural algorithm in \lstinline|rpp| to implement~\eqref{align: cp analitically}. As expected, the input pair $(x,y)$ is part of \lstinline|cp_in| output. The suffix ``\lstinline|in|'' in the name ``recalls'' exactly this aspect. In order to drop $(x,y)$ from the output of \lstinline|cp_in|, and obtain \lstinline|cp| as in Figure~\ref{sfig:cp}, applying {\it Bennet's trick}, we need \lstinline|cu_in⁻¹|, i.e. the inverse of \lstinline|cu_in| which is new, as compared to \cite{DBLP:journals/tcs/PaoliniPR20}. The intuition behind \lstinline|cu_in| is as follows. Let us fix any point $ (x, y) \in \NN^2 $. We can realize that, starting from the origin, if we follow as many steps as the value $ \rppcp(x, y) $ in Figure~\ref{sfig:cantor as a path}, we stop exactly at $ (x,y) $. The function, expressed in standard functional notation, that, given the current point $ (x,y) $, identifies the next one to move to in the path of Figure~\ref{sfig:cantor as a path} is:
\begin{align*}
%    \label{align:next function}
    \operatorname{step}(x,y) & =
    \begin{cases} (x+1,y-1) &  y > 0 \\ (0, x+1) &   y = 0 \end{cases}
    \enspace .
\end{align*}
We implement $ \operatorname{step}(x,y) $ in \lstinline|rpp| as \lstinline|step[Su;;Sw]|. Figures~\ref{sfig:cu step y > 0}, and~\ref{sfig:cu step y = 0} represent two runs of \lstinline|step[Su;;Sw]| to give visual evidence that \lstinline|step[Su;;Sw]| implements $\operatorname{step}(x,y)$. Colored occurrences of $ y $ show the relevant part of the computational flow. Note that we cannot implement $ \operatorname{step}(x,y) $ by using the conditional \lstinline|It| directly on $y$, because in the computation we also want to modify the value of $y$. Finally, as soon as we get \lstinline|cu_in| by iterating \lstinline|step[Su;;Sw]| as in Figure~\ref{sfig:cu input}, we can define \lstinline|cp| (Figure~\ref{sfig:cp}), and \lstinline|cu| (Figure~\ref{sfig:cu}).  \qed

\begin{figure}
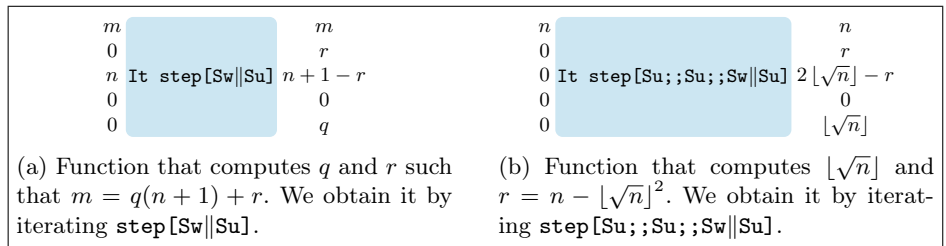

    \begin{subfigure}{.475\linewidth}
        \centering
\scalebox{.85}
{        $ \begin{NiceMatrix}
            m & \bloch{5-1}{\mbox{\lstinline|It step[Sw‖Su]|}} & m         \\
            0 &                      & r         \\
            n &                      & n + 1 - r \\
            0 &                      & 0         \\
            0 &                      & q         \\
        \end{NiceMatrix} $
}        \caption{Function that computes $ q $ and $ r $ such that $m = q(n + 1) + r$. We obtain it by iterating \lstinline|step[Sw‖Su]|.}
        \label{sfig:div}
    \end{subfigure}
    \hfill
    \begin{subfigure}{.475\linewidth}
        \centering
 \scalebox{.85}
 {       $\begin{NiceMatrix}
            n & \bloch{5-1}{\mbox{\lstinline|It step[Su;;Su;;Sw‖Su]|}} & n      \\
            0 &                       & r                      \\
            0 &                       & 2 \floor{\sqrt{n}} - r \\
            0 &                       & 0                      \\
            0 &                       & \floor{\sqrt{n}}       \\
        \end{NiceMatrix}
        $
}
        \caption{Function that computes $ \floor{\sqrt{n}}$ and $ r = n - \floor{\sqrt{n}}^2 $. We obtain it by iterating \lstinline|step[Su;;Su;;Sw‖Su]|.}
        \label{sfig:sqrt}
    \end{subfigure}%
\caption{Quotient/Reminder and Square root.}
\end{figure}

%=============================
\paragraph{Quotient and reminder.}
%In defining the function $\rppcui$ we framed it as a "path" in a two-dimensional grid.
Let us focus on the path in Figure~\ref{sfig:Quot./Rem.}.
It starts at $(0,n)$ (with $ n = 4 $), and, at every step, the next point is in \emph{direction} $(+1,-1)$. When it reaches $(n,0)$ (with $ n = 4 $), instead of jumping to $(0,n+1)$, as in Figure~\ref{sfig:cantor as a path}, it lands again on $(0,n)$. The idea is to keep looping on the same diagonal. This behavior can be achieved by iterating \lstinline|step[Sw‖Su]|. Figure~\ref{sfig:div} shows that we are doing modular arithmetic.
Globally, it takes $n+1$ steps from $ (0,n) $ to itself by means of \lstinline|step[Sw‖Su]|.
Specifically, if we assume we have performed $m$ steps along the diagonal, and we are at point $ (x,y) $, we have that $x \equiv m \pmod{n+1}$ and $0 \le x \le n$.
So, if we increase a counter by one each time we get back to $(0,n)$ we can calculate quotient and reminder.
\qed

%====================
\paragraph{Truncated Square root.}
Let us focus on the path in Figure~\ref{sfig:Square root}.
It starts at $(0,0)$. Whenever it reaches $(x,0)$ it jumps to $(0,x+2)$, otherwise the next point is in \emph{direction} $(+1,-1)$.
The behavior can be achieved by iterating \lstinline|step[Su;;Su;;Sw‖Su]| as in Figure~\ref{sfig:sqrt}.
In order to compute $ \floor{\sqrt{n}} $, besides implementing the above path, the function \lstinline|step[Su;;Su;;Sw‖Su]| counts in $ k $ the number of jumps occurred so far along the path. In particular, starting from $ (0,0) $, the first jump occurs in the first step; the next one in the $(1+3)$th, then the $(1+3+5)$th, then the $(1+3+5+7)$th etc. Since we know that $1 + 3 + \dots + (2k - 1) = k^2$ for any $k$, letting $n$ be the number of iterations (and hence the numbers of steps) we have that $k$ is such that $k^2 \le n < (k+1)^2$; i.e. $k = \floor{\sqrt{n}}$. \qed

\begin{remark}
The value $2 \floor{\sqrt{n}} - r$ can be canceled out by adding $r$, and subtracting $\floor{\sqrt{n}}$ twice.
What we \emph{cannot} eliminate is the ``remainder'' $r = n - \left\lfloor \sqrt{n} \right\rfloor^2$ because the \emph{function} Square root
cannot be inverted in $ \ZZ $, and the algorithm cannot forget it.
\qeda
\end{remark}

\paragraph{The $\rppmkpair$ function.}
Figure~\ref{sfig:mkpair} shows the behavior of the function $\rppmkpair$.
It is very similar to the one of $\rppcp$, but it uses an alternative algorithm described in \cite{Carneiro19}.
Here we do not describe it in detail because it's just a composition of sums, products and square roots,
which have already been discussed.

\paragraph{A note on the mechanization of proofs}
We recall once more that everything defined here above has been proved correct in \LEAN.
For example, in \cite{MalettoRPPLEAN2021}, one can define as we did the \lstinline|rpp| term \lstinline|sqrt| and prove its behavior:
\begin{lstlisting}
lemma sqrt_def (n : ℕ) (X : list ℤ) :
 ‹sqrt›(n::0::0::0::0::X) =
           n::(n-√n*√n)::(√n+√n-(n-√n*√n))::0::√n::X
\end{lstlisting}
In order to prove these theorems we make use of a {\it tactic} (which is a command used to build proofs)
known as \lstinline|simp|, which is able to automatically simplify expressions
until one gets a trivial identity. What is meant by simplify, is that theorems which state an equality like $LHS = RHS$ (e.g. \lstinline|sqrt_def|)
can be marked with the attribute \lstinline|@[simp]|, which means that everytime the \lstinline|simp| tactic is invoked in another proof,
if the equality to be proved has the expression $LHS$ then it will be substituted with $RHS$, often making it simpler.

This technique is really powerful, because it makes it possible to essentially automate many proofs of theorems which in turn can be marked with \lstinline|@[simp]|
and be used to prove yet more theorems.

%=====================
\section{Proving in \LEAN that \RPP is \PRF-complete}
\label{section:The UPRF-completeness of RPP}
We formally show in \LEAN that the class of functions we can express as (algorithms) in \lstinline|rpp| contains at least \PRF; so, we say ``\lstinline|rpp| is \PRF-complete''. The definition of \PRF that we take as reference is one of the two available in \MATHLIB library of \LEAN. Once recalled and commented it briefly, we shall proceed with the main aspects of the \PRF-completeness of \lstinline|rpp|.

\begin{figure}
\begin{lstlisting}[basicstyle=\small]
inductive primrec:(ℕ → ℕ) → Prop
| zero: primrec (λ (n:ℕ), 0)
| succ: primrec succ
| left: primrec (λ (n:ℕ), (unpair n).fst)
| right: primrec (λ (n:ℕ), (unpair n).snd)
| pair {F G}: primrec F → primrec G → primrec (λ (n:ℕ), mkpair (F n) (G n))
| comp {F G}: primrec F → primrec G → primrec (λ (n:ℕ), F (G n))
| prec {F G}: primrec F → primrec G → primrec
(unpaired (λ (z n:ℕ), nat.rec (F z) (λ (y IH:ℕ), G (mkpair z (mkpair y IH))) n))
\end{lstlisting}
\caption{\lstinline|primrec| defines \PRF in \MATHLIB of \LEAN.}
\label{fig: primrec}
\end{figure}

%=====================
\subsection{Primitive Recursive Functions {\normalfont \lstinline|primrec|} of \MATHLIB }
\label{section:Unary Primitive Recursive Functions}

Figure~\ref{fig: primrec} recalls the definition of \PRF from \cite{Carneiro-primrecMathlib} available in \MATHLIB that we take as reference. It is an inductively defined \lstinline|Prop|osition \lstinline|primrec| that requires a \emph{unary} function with type \lstinline|ℕ → ℕ| as argument. Specifically, \lstinline|primrec| is the least collection of functions \lstinline|ℕ → ℕ| with a given set of base elements, closed under some composition schemes.

\paragraph{Base functions.}
The \emph{constant} function \lstinline|zero| yields \lstinline|0| on every of its inputs.
The \lstinline|succ|\emph{essor} gives the natural number next to the one taken as input.
The two \emph{projections} \lstinline|left|, and \lstinline|right| take an argument \lstinline|n|, and extract a left, or a right, component from it as \lstinline|n| was the result of pairing two values \lstinline|x,y:ℕ|. The functions that \lstinline|primrec| relies on to encode/decode pairs on natural numbers as a single natural one are \lstinline|mkpair:ℕ → ℕ → ℕ|, and \lstinline|unpair:ℕ → ℕ × ℕ|. The first one builds the value \lstinline|mkpair x y|, i.e. the number of steps from the origin to reach the point with coordinates \lstinline|(x,y)| in the path of Figure~\ref{sfig:mkpair}. The function \lstinline|unpair:ℕ → ℕ × ℕ| takes the number of steps to perform on the same path. Once it stops, the coordinates of that point are the two natural numbers we are looking for. So, \lstinline|mkpair|/\lstinline|unpair| are an alternative to Cantor Pairing/Un-pairing.

\paragraph{Composition schemes.}
Three schemes exist in \lstinline|primrec|, each depending on parameters \lstinline|f,g:primrec|.
The scheme \lstinline|pair| builds the function that, taken a value \lstinline|n:ℕ|, gives the unique value in \lstinline|ℕ| that encodes the pair of values \lstinline|F n|, and \lstinline|G n|; everything we might pack up by means of \lstinline|pair|, we can unpack with \lstinline|left|, and \lstinline|right|.

The scheme \lstinline|comp| composes \lstinline|F,G:primrec|.

The \emph{primitive recursion} scheme \lstinline|prec| can be ``unfolded'' to understand how it works; this reading will ease the description of how to encode it in \lstinline|rpp|.
Let \lstinline|F|, \lstinline|G| be two elements of \lstinline|primrec|. We see \lstinline|prec| as encoding the function:
\begin{align}
\label{align:function H}
H[\mbox{\lstinline|F|},\mbox{\lstinline|G|}](x) & =
R[\mbox{\lstinline|G|}]\big( \mbox{\lstinline|F|} \big((x)_1 \big) , (x)_2 \big)
\end{align}
where:
(i) $(x)_1$ denotes \lstinline|(unpair| x\lstinline|).fst|,
(ii) $(x)_2$ denotes \lstinline|(unpair| x\lstinline|).snd|,
and
(iii) $R [\mbox{\lstinline|G|}] $ behaves as follows:
\begin{equation}
\label{equation:function R}
\begin{aligned}
    R[\mbox{\lstinline|G|}](z,0) & = z \\
    R[\mbox{\lstinline|G|}](z,n+1) &
    = \mbox{\lstinline|G|} \big(
      \gl z, \gl n, R[\mbox{\lstinline|G|}](z,n) \gr \gr
      \big)
   \enspace ,
\end{aligned}
\end{equation}
defined using the built-in recursive scheme \lstinline|nat.rec| on \lstinline|ℕ|, and $ \gl a, b \gr $ denotes $ (\mbox{\lstinline|mkpair|}\ a\ b) $.

%=================
\subsection{The main point of the proof}
In order to formally state what we mean for \lstinline|rpp| to be \PRF-complete, in \LEAN we need to say when, given \lstinline|F:ℕ → ℕ|, we can \emph{encode} it by means of some \lstinline|f:rpp|:
\begin{lstlisting}
  def encode (F:ℕ → ℕ) (f:rpp) :=
     ∀ (z:ℤ) (n:ℕ), ‹f› (z::n::repeat 0 (f.arity-2))
                            = (z+(F n))::n::repeat 0 (f.arity-2)
\end{lstlisting}
says that, fixed \lstinline|F:ℕ → ℕ|, and \lstinline|f|, the statement \lstinline|(encode F f)| holds if the evaluation of \lstinline|‹f›|, applied to any argument \lstinline|(z::n::0::...::0)|, with as many occurrences of trailing \lstinline|0|s as \lstinline|f.arity-2|, gives a list with form \lstinline|((z+(F n))::n::0::...::0)| such that: (i) the first element is the original value \lstinline|z| increased with the result \lstinline|(F n)| of the function we want to encode; (ii) the second element is the initial \lstinline|n|; (iii) the trailing \lstinline|0|s are again as many as \lstinline|f.arity-2|. In \LEAN we can prove:
\begin{lstlisting}
theorem completeness (F:ℕ → ℕ): primrec F → ∃ f:rpp, encode F f
\end{lstlisting}
which says that we know how to build \lstinline|f:rpp| which \lstinline|encode|s \lstinline|F|, for every well formed \lstinline|F:ℕ → ℕ|, i.e. such that \lstinline|primrec F| holds.

The proof proceeds by induction on the proposition \lstinline|primrec|, which generates 7 sub-goals. We illustrate the main arguments to conclude the most interesting case which requires to encode the composition scheme \lstinline|prec|.

\begin{remark}
Many aspects we here detail out were simply missing in the original \PRF-completeness proof for \RPP in \cite{DBLP:journals/tcs/PaoliniPR20}. \qeda
\end{remark}

The inductive hypothesis to show that we can encode \lstinline|prec| is that, for any given \lstinline|F,G:ℕ → ℕ| such that \lstinline|(primrec F):Prop|, and \lstinline|(primrec G):Prop|, both \lstinline|f,g:rpp| exist such that \lstinline|(encode F f)|, and \lstinline|(encode G g)| hold. This means that
$ \mbox{\lstinline|f|}\ (z\mbox{\lstinline|::|}n\mbox{\lstinline|::|}\boldsymbol{0}) =
(z+\mbox{\lstinline|F|}\ n)\mbox{\lstinline|::|}n\mbox{\lstinline|::|}\boldsymbol{0} $, and
$ \mbox{\lstinline|g|}\ z\mbox{\lstinline|::|}n\mbox{\lstinline|::|}\boldsymbol{0} =
(z+\mbox{\lstinline|G|}\ n)\mbox{\lstinline|::|}n\mbox{\lstinline|::|}\boldsymbol{0} $, where $ \boldsymbol{0} $ stands for a sufficiently long list of $ 0 $s.

\begin{figure}
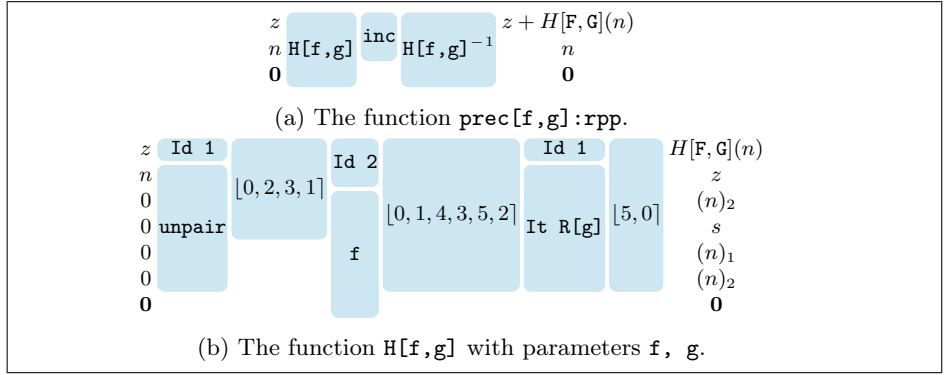

\newcommand{\bz}{\boldsymbol{0}}
\newcommand{\by}{\boldsymbol{y}}
\newcommand{\bid}{\bloch{1-1}{\mbox{\lstinline|Id 1|}}}
\newcommand{\bidt}{\bloch{2-1}{\mbox{\lstinline|Id 2|}}}
\newcommand{\bsw}{\bloch{2-1}{\mbox{\lstinline|Sw|}}}
\newcommand{\bpi}{\bloch{3-1}{\mbox{\lstinline|H[f,g]|}}}
\newcommand{\binc}{\bloch{2-1}{\mbox{\lstinline|inc|}}}
\newcommand{\bup}{\bloch{5-1}{\mbox{\lstinline|unpair|}}}
\newcommand{\bpiinv}{\bloch{3-1}{\mbox{\lstinline|H[f,g]⁻¹|}}}
\newcommand{\bff}{\bloch{5-1}{\mbox{\lstinline|f|}}}
\newcommand{\bfg}{\bloch{3-1}{\mbox{\lstinline|g|}}}
\newcommand{\bre}{\bloch{4-1}{\rpprewire{0,2,3,1}}}
\newcommand{\brs}{\bloch{6-1}{\rpprewire{0,1,4,3,5,2}}}
\newcommand{\brc}{\bloch{6-1}{\rpprewire{5,0}}}
\newcommand{\bipsg}{\bloch{5-1}{\mbox{\lstinline|It R[g]|}}}
\begin{subfigure}{.95\textwidth}
\centering
\scalebox{.9}
{$\begin{NiceMatrix}
 z  &\bpi&\binc&\bpiinv&z + H[\mbox{\lstinline|F|},\mbox{\lstinline|G|}] (n)\\
 n  &    &    &       &n\\
 \bz&    &    &       &\bz
 \end{NiceMatrix}$}
\caption{The function \lstinline|prec[f,g]:rpp|.}
\label{sfig:prec[f,g]}
\end{subfigure}
\\
\begin{subfigure}{.95\textwidth}
\centering
\scalebox{.9}{
$\begin{NiceMatrix}
 z  &\bid&\bre&\bidt&\brs&\bid  &\brc& H[\mbox{\lstinline|F|},\mbox{\lstinline|G|}] (n)  \\
 n  &\bup&    &     &    &\bipsg&    & z      \\
 0  &    &    &\bff &    &      &    & (n)_2  \\
 0  &    &    &     &    &      &    & s      \\
 0  &    &    &     &    &      &    & (n)_1  \\
 0  &    &    &     &    &      &    & (n)_2  \\
 \bz&    &    &     &    &      &    & \bz
\end{NiceMatrix}$
 }
\caption{The function \lstinline|H[f,g]| with parameters \lstinline|f, g|.}
\label{sfig:precfwd[f,g]}
\end{subfigure}
\caption{Encoding \lstinline|prec| of Figure~\ref{fig: primrec} in \lstinline|rpp|.}
\label{fig:composing encode prec}
\end{figure}

Figure~\ref{sfig:prec[f,g]}, where we assume $ z = 0 $, defines \lstinline|prec[f,g]:rpp| such that \lstinline|(encode (prec F G) prec[f,g]):Prop| holds, and \lstinline|H[f,g]| encodes
$ H[\mbox{\lstinline|F|},\mbox{\lstinline|G|}]$ as in~\eqref{align:function H}.
The term \lstinline|It R[g]| in \lstinline|H[f,g]| encodes~\eqref{equation:function R} by iterating \lstinline|R[g]| from the initial value given by \lstinline|f|.

Figure~\ref{fig:main step RPF-completeness} splits the definition of \lstinline|R[g]| into three logical parts.
Figure~\ref{sfig:encode prec 1st step} packs everything up by means of \lstinline|mkpair| to build the argument $ R[\mbox{\lstinline|G|}](z,n) $ of \lstinline|g|; by induction we get
$ R[\mbox{\lstinline|G|}](z,n+1) $.
In Figure~\ref{sfig:encode prec 2nd step}, \lstinline|unpair| unpacks $ \gl z, \gl n, R[\mbox{\lstinline|G|}](z,n) \gr \gr $ to expose its component to the last part.
Figure~\ref{sfig:encode prec 3rd step} both increments $ n $, and packs $ R[\mbox{\lstinline|G|}](z,n) $ into $ s $, by means of \lstinline|mkpair|, because $ R[\mbox{\lstinline|G|}](z,n) $ has become useless once obtained $ R[\mbox{\lstinline|G|}](z,n+1) $ from it. Packing $ R[\mbox{\lstinline|G|}](z,n) $ into $ s $, so that we can eventually recover it, \emph{is mandatory}. We cannot ``replace'' $ R[\mbox{\lstinline|G|}](z,n) $ with $ 0 $ because that would not be a reversible action.

\begin{remark}
The function \lstinline|cp| in Figure~\ref{sfig:cp} can replace \lstinline|mkpair| in Figure~\ref{sfig:encode prec 3rd step} as a bijective map $ \NN^2$ into $\NN$. Indeed, the original \PRF-completeness of \RPP relies on \lstinline|cp|. We favor \lstinline|mkpair| to take the most out of \MATHLIB. \qeda
\end{remark}

\begin{figure}
    \begin{subfigure}{.975\textwidth}
        \centering
        \scalebox{.9}{
            $\begin{NiceMatrix}
                s&\bloch{2-1}{\mbox{\lstinline|Id 2|}}&\bloch{1-1}{\mbox{\lstinline|Id 1|}}& s &\bloch{1-1}{\mbox{\lstinline|Id 1|}}&\bloch{1-1}{\mbox{\lstinline|Id 1|}}& s
                \\
                z&& \bloch{5-1}{\mbox{\lstinline|mkpair|}} & \gl z, \gl n, R[\mbox{\lstinline|G|}](z,n) \gr \gr & \bloch{2-1}{\mbox{\lstinline|Sw|}} & \bloch{5-1}{\mbox{\lstinline|g|}}&R[\mbox{\lstinline|G|}](z,n+1)
                \\
                n& \bloch{4-1}{\mbox{\lstinline|mkpair|}} &&0&&& \gl z, \gl n, R[\mbox{\lstinline|G|}](z,n) \gr \gr
                \\
                R[\mbox{\lstinline|G|}](z,n) &                         &                         & 0                                       &                     &                & 0                                       \\
                0                 &                         &                         & 0                                       &                     &                & 0                                       \\
                \boldsymbol{0}    &                         &                         & \boldsymbol{0}                          &                     &                & \boldsymbol{0}                          \\
            \end{NiceMatrix}$
        }
        \caption{Build the argument $ \gl z, \gl n, R[\mbox{\lstinline|G|}](z,n) \gr \gr $ of \lstinline|g|.}
        \label{sfig:encode prec 1st step}
    \end{subfigure}
    \\
    \begin{subfigure}{.975\textwidth}
        \centering
        \scalebox{.9}{
            $   \begin{NiceMatrix}
                s &\bloch{2-1}{\mbox{\lstinline|Id 2|}}&\bloch{3-1}{\mbox{\lstinline|Id 3|}}&s& \bloch{5-1}{\rpprewire{2,3,1,0,4}} & z                   \\
                R[\mbox{\lstinline|G|}](z,n+1)                     &                         &                         & R[\mbox{\lstinline|G|}](z,n+1) &                            & n                   \\
                \gl z, \gl n, R[\mbox{\lstinline|G|}](z,n) \gr \gr & \bloch{4-1}{\mbox{\lstinline|unpair|}} &                         & z                   &                            & R[\mbox{\lstinline|G|}](z,n+1) \\
                0                                       &                         & \bloch{3-1}{\mbox{\lstinline|unpair|}} & n                   &                            & s                   \\
                0                                       &                         &                         & R[\mbox{\lstinline|G|}](z,n)   &                            & R[\mbox{\lstinline|G|}](z,n)   \\
                \boldsymbol{0}                          &                         &                         & \boldsymbol{0}      &                            & \boldsymbol{0}      \\
            \end{NiceMatrix}$
        }
        \caption{Unpack $ \gl z, \gl n, R[\mbox{\lstinline|G|}](z,n) \gr \gr $ to let its elements available.}
        \label{sfig:encode prec 2nd step}
    \end{subfigure}
    \\
    \begin{subfigure}{.975\textwidth}
        \centering
        \scalebox{.9}{
            $    \begin{NiceMatrix}
                z&\bloch{1-1}{\mbox{\lstinline|Id 1|}}& z& \bloch{4-1}{\rpprewire{3,0,1,2}} & s'                  \\
                n                   & \bloch{1-1}{\mbox{\lstinline|Su|}}     & n + 1               &                         & z                   \\
                R[\mbox{\lstinline|G|}](z,n+1) &\bloch{1-1}{\mbox{\lstinline|Id 1|}} & R[\mbox{\lstinline|G|}](z,n+1) &                         & n + 1               \\
                s                   & \bloch{3-1}{\mbox{\lstinline|mkpair|}} & s'                  &                         & R[\mbox{\lstinline|G|}](z,n+1) \\
                R[\mbox{\lstinline|G|}](z,n)   &                         & 0                   &                         & 0                   \\
                \boldsymbol{0}      &                         & \boldsymbol{0}      &                         & \boldsymbol{0}      \\
            \end{NiceMatrix}$
        }
        \caption{Increment $ n $, and store $ R[\mbox{\lstinline|G|}](z,n) $ to keep the whole process reversible.}
        \label{sfig:encode prec 3rd step}
    \end{subfigure}
    \caption{Encoding $ R[\mbox{\lstinline|G|}] $ in~\eqref{equation:function R} as \lstinline|R[g]:rpp|.}
    \label{fig:main step RPF-completeness}
\end{figure}

%=====================
\section{Conclusion and developments}
\label{section:Conclusion and developments}
We give a concrete example of reversible programming in a proof-assistant. We think it is a valuable operation because programming reversible algorithms is not as much wide-spread as classical iterative/recursive programming, in particular by means of a tool that allows us to certify the result.
Other proof assistants have been considered, and in fact the same theorems have also been proved in \COQ, but we found that the use of the \MATHLIB library together with the \lstinline|simp| tactic made our experience with \LEAN much smoother.
Furthermore, our work can migrate to \LEANFour whose stable release is announced in the near future. \LEANFour exports its source code as efficient \CPP code \cite{2021-LEAN4-MouraUllrich}; our and other reversible algorithms can become efficient extensions of \LEANFour, or standalone, and \CPP applications.

The most application-oriented obvious goal to mention is to keep developing a Reversible Computa\-tion-centered certified software stack, spanning from a programming formalism more friendly than \lstinline|rpp|, down to a certified emulator of \PISA, passing through compilator, and optimizer whose properties we can certify. For example, we can also think of endowing \PISA emulators with energy-consumption models linked to the entropy that characterize the reversible algorithms we program, or the \PISA object code we can generate from them.

A more speculative direction, is to keep exploring the existence of programming schemes in \lstinline|rpp| able to generate functions, other than Cantor Pairing, etc., which we can see as discrete space-filling functions, whose behavior we can describe as steps, which we count, along a path in some space.

%------------------
\bibliographystyle{abbrv}
\bibliography{bib-minimal}

%\appendix

\end{document}